\documentclass[notitlepage,12pt]{article}%
\usepackage{amsfonts}
\usepackage{amssymb}
\usepackage{amsmath}
\usepackage{amsthm}
\usepackage[normalem]{ulem}
\usepackage{multirow}
\usepackage[flushleft]{threeparttable}
\usepackage{array}
\usepackage{setspace}
\usepackage{ragged2e}
\usepackage{lscape}
\usepackage{graphicx}
\graphicspath{ {./images/} }
\usepackage{amsfonts}
\usepackage{scalefnt}
\usepackage{natbib}

\providecommand{\U}[1]{\protect\rule{.1in}{.1in}}
\advance \topmargin by -\headheight
\advance \topmargin by -\headsep
\evensidemargin \oddsidemargin
\newtheorem{theorem}{Theorem}

\DeclareMathOperator*{\plim}{plim}

\input{tcilatex}
\begin{document}
\thispagestyle{empty}

\title{Another Look at the Linear Probability Model and Nonlinear Index
Models}
\author{Kaicheng Chen\thanks{Department of Economics, Michigan State University. Email: chenka19@msu.edu}
\and Robert S. Martin\thanks{Division of Price and Index Number Research, Bureau of Labor Statistics. Email: Martin.Robert@bls.gov}
\and Jeffrey M. Wooldridge\thanks{{Department of Economics, Michigan State University. Email:wooldri1@msu.edu}}}
\date{\today}
\maketitle

\begin{abstract}
We reassess the use of linear models to approximate response probabilities of binary outcomes, focusing on average partial effects (APE). We confirm that linear projection parameters coincide with APEs in certain scenarios. Through simulations, we identify other cases where OLS does or does not approximate APEs and find that having large fraction of fitted values in $[0,1]$ is neither necessary nor sufficient. We also show nonlinear least squares estimation of the ramp model is consistent and asymptotically normal and is equivalent to using OLS on an iteratively trimmed sample to reduce bias. Our findings offer practical guidance for empirical research.
\end{abstract}

\bigskip


\bigskip


\bigskip

\setcounter{page}{0} \thispagestyle{empty} 
\pagestyle{plain}

\section{Introduction}

When an outcome variable, $y$, is binary, empirical researchers usually
choose between two general strategies given a vector of (exogenous)
explanatory variables, $\mathbf{x}$: (i) approximate the response
probability, $P\left( y=1|\mathbf{x}\right) $, using a model linear in
parameters or (ii) use a nonlinear model, such as logit or probit. The first
strategy is commonly known as using a \textit{linear probability model}
(LPM). The benefits of the LPM are well-known and include ease of
interpretation, simple estimation, and straightforward extension to
situations with endogenous explanatory variables (so that instrumental
variables are used) and panel data settings with unobserved heterogeneity.
The shortcomings of the LPM are also well known, and discussed in most
introductory econometrics texts; see, for example, \citealp[Section 7.5]{wooldridge2019introductory}. More advanced discussions of the LPM recognize that one should not take the
linear model for $P\left( y=1|\mathbf{x}\right) $ literally but only as an
approximation. The approximation can be exact in special cases---such as
when $\mathbf{x}$ consists of binary indicators that are exhaustive and
mutually exclusive---and it may be poor in other cases. However, for the
most part, prediction is not the primary use of LPMs specifically or binary
response models generally. Rather, researchers are largely interested in
using binary response models to measure ceteris paribus or causal effects,
and it is from this perspective that the LPM approximation should be
evaluated. \citealp[Section 3.4.1]{angrist2009mostly} and \citealp[Section 15.2]{wooldridge2010econometric} take this perspective. \citealp[Section 15.6, p. 579]{wooldridge2010econometric} shows how the
results of \cite{Stoker1986} can be applied to OLS estimation of the parameters
in a linear probability model. Remarkably, there are situations where the
linear projection exactly recovers the APEs across a broad range of binary
response models. As is well known---see, for example, \citealp[Chapter 4]{wooldridge2010econometric}---under standard sampling assumptions---OLS consistently
estimates the parameters of the linear projection (LP).

Even though it is natural to study the LPM from the linear projection
perspective, this opinion is not universally held. In an influential paper,
\cite{Horrace2006} study both the bias and inconsistency of the OLS
estimator for the parameters of an underlying piecewise linear model for the
response probability that ensures the probabilities are in the unit
interval.\footnote{\cite{Horrace2006} defines the LPM as the piecewise linear ramp model. However, in this paper we differentiate between the “ramp model” and the “LPM” (which is linear everywhere).} The Horrace-Oaxaca (H-O) paper is regularly cited in empirical
research, sometimes as a cautionary tale in using the LPM and sometimes as
support for using the LPM when relatively few fitted values lie outside the
unit interval. (In the previous two years, H-O has almost 200 Google Scholar
citations.) While H-O take the piecewise linear model seriously, much if not most of the citing literature seeks to use their results to choose between the LPM and an alternative like probit or logit.\footnote{See, for example, Footnote 20 of \cite{van2022effects}.}

In the current paper, we revisit the H-O framework but, rather than focus on
parameters, we focus on APEs. We show that H-O set up the problem so that,
in general, the response probability is nonlinear in the underlying linear
index, $\mathbf{x\beta }=\beta _{1}+\beta _{2}x_{2}+\cdots +\beta _{K}x_{K}$%
. The nonlinear function of $\mathbf{x\beta }$---sometimes called the ramp function---is piecewise linear and
continuous, but it is not strictly increasing, and it is nondifferentiable at
two inflection points. Nevertheless, under fairly weak assumptions, one can
define the average partial effects, and these are necessarily smaller in
magnitude than the corresponding parameter in the underlying nonlinear
model. Consequently, H-O's focus on parameters rather than APEs is
essentially the same as focusing on parameters in smooth response
probabilities such as the logit and probit functions. Therefore, any
conclusions about the usefulness of the LPM should be reexamined from the
perspective of identifying APEs rather than coefficients. 

It is important to understand that we are not necessarily advocating the H-O ramp
function as an especially sensible model of the response probability.
Rather, we primarily study that specification from the perspective of average partial
effects to determine how the OLS estimator holds up. Briefly, in some cases
OLS does a very good job of approximating the APEs even when a larger
percentage of the fitted values are outside the unit interval.\ Conversely,
in other cases, OLS does a very poor job of approximating the APEs even
when a high percentage of the fitted values are within the unit interval. A
practical implication is that there is little justification for how the H-O
study is cited in empirical research.

We compare OLS to a few nonlinear competitors, including probit and logit quasi-maximum likelihood estimation (QMLE), as natural benchmarks. H-O cite a few theoretical rationalizations for the ramp model, so it also makes sense to see if a consistent estimator exists which takes it seriously. H-O suggest trimming the sample of fitted values outside the unit interval and re-estimating using OLS, but do not present any theoretical or simulation results.\footnote{In unreported simulations, we found that trimming the sample once did not necessarily improve performance over OLS for estimating the APEs.} In Section 4, we show that nonlinear least squares estimation (NLS) using the ramp function is consistent and asymptotically normal under mild assumptions. We also give a variance estimator for the asymptotic variance of the NLS estimator and provide a consistency result for this. We are unaware of any other studies using NLS to estimate the ramp model. We also examine an iterative trimming OLS (ITO) procedure similar in spirit to H-O's suggestion and show it is equivalent to numerically minimizing the NLS objective function using the well-known Newton-Raphson algorithm. In simulations, for estimating the APEs, we find that NLS estimation of the ramp function performs comparably to quasi-MLE estimation of the logit and probit models and has good finite sample properties even when OLS estimation of the LPM does not. For completeness, we also consider a local linear estimation of a nonparametric model of the conditional mean, but it seems for our data generating processes there is not much gain in using a nonprarametric model compared to other nonlinear models. 

In Section 2 we present the population model equivalent of the Horrace-Oaxaca model and show that it is equivalent to a latent variable
model with uniformly distributed errors. We\ also derive the average partial
effects of the so-called \textquotedblleft ramp function.\textquotedblright\
In Section 3 we extend the discussion in \citealp[Section 15.6]{wooldridge2010econometric}
and show that, when the covariates have a multivariate normal distribution,
the linear projection identifies the APEs. Section 5 contains several
simulations comparing sample APEs produced by OLS, NLS, Probit QMLE, Logit QMLE, and Local Linear estimations when the true response probability follows the ramp model. We provide some results not covered by the existing theories and show that a large fraction of fitted values in $[0,1]$ is neither sufficient or necessary condition for OLS to well-approximate the APEs. 

In Section 6, we apply LPM, ramp, probit, and logit models to an empirical study of discrimination in mortgage lending decisions. The LPM estimated by OLS estimation, with a full set of interactions between the race indicator and the control
variables, delivers a notably smaller and marginally statistically
significant estimate of the race effect on the approval probability. The NLS, probit QMLE, and logit QMLE are very similar and all statistically
significant at the 0.2\% level---both because the estimated effects are
larger but also because the (robust) standard errors are notably smaller. In Section 7, we conclude with some implications for empirical research.

\section{The Ramp Model and the Linear Projection}

One of the key features of the Horrace-Oaxaca setting is that it imposes the
logical bounds on the response probability in a setting where, over some of
its range, the response probability is linear in an index. Almost all of the
important arguments are in terms of the underlying population model, so that
is our focus. Let $y$ be the binary outcome variable and $\mathbf{x}$ the $%
1\times K$ vector of explanatory vairables, where $x_{1}\equiv 1$ allows for
an intercept in the index. Defining $p(\mathbf{x}) = P(y=1 | \mathbf{x})$, then H-O's specification can be written as $ p\left( \mathbf{x}\right) = R(\mathbf{x\beta })$ where 
where $\mathbf{x\beta }=\beta _{1}+\beta _{2}x_{2}+\cdots +\beta _{K}x_{K} $ is the linear index and  \vspace{-3mm}
\begin{align}
R\left( z\right) = \begin{cases}
            0, & \text{$\mathbf{z}\leq 0$}\\
            \mathbf{z}, & \text{$\mathbf{z}\in(0,1)$}\\
            1, & \text{$\mathbf{z}\geq 1$}
            \end{cases} \tag{2.1} 
            \label{HOspec} \vspace{-3mm}
\end{align}
is the \textquotedblleft ramp function\textquotedblright. This response probability was also suggested by \cite{horowitz2001binary} as
being suitable when one starts with a linear model for $p\left( \mathbf{x}%
\right) $ but wants to ensure that the probabilities are within the unit
interval. The ramp function, which is piecewise linear, is plotted in Figure 1.
\begin{center}
    \includegraphics[scale=0.4]{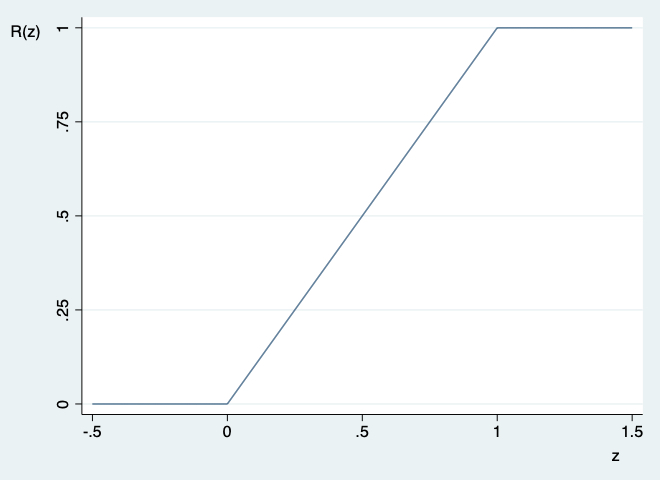} \\
    \textbf{Figure 1}: The Ramp Function
\end{center}

In defining partial effects we must recognize that the function $R\left(
z\right) $ is nondifferentiable at $z=0$, $z=1$. A very common assumption
imposed in the semiparametric literature on binary response models is that
at least one element of $\mathbf{x}$ is continuous, and that element has a
nonzero coefficient. Then $\mathbf{x\beta }$ is continuous, and so  \vspace{-3mm}
\begin{equation*}
P\left( \mathbf{x\beta }=0\right) =P\left( \mathbf{x\beta }=1\right) =0.  \vspace{-3mm}
\end{equation*}
In what follows, we\ maintain that $\mathbf{x\beta }$ is continuous so that
partial effects are well-defined with probability one.

Now let $x_{j}$ be a continuously distributed explanatory variable. For
simplicity, the discussion here assumes that $x_{j}$ appears only by itself.
If the model includes quadratics, interactions, and so on then the details
become more complicated but the conclusions do not change substantively.

We can define a partial effect function as the derivative of $R\left( 
\mathbf{x\beta }\right) $ and ignore points where the derivative does not
exist:  \vspace{-3mm}
\begin{equation*}
PE_{j}\left( \mathbf{x}\right) =\frac{\partial p}{\partial x_{j}}\left( 
\mathbf{x}\right) =\beta _{j}1\left[ 0\leq \mathbf{x\beta }\leq 1\right]  \vspace{-3mm},
\end{equation*}%
where $1\left[ \cdot \right] $ is the indicator function. This expression
for $PE_{j}\left( \mathbf{x}\right) $ uses the fact that  \vspace{-3mm}
\begin{eqnarray*}
\frac{dR}{dz}\left( z\right) &=&1\text{, }z\in \left( 0,1\right) \\
&=&0\text{ if }z<0\text{ or }z>1, \vspace{-3mm}
\end{eqnarray*}
and is undefined at $z=0$ or $z=1$. In obtaining the average partial effect
of $x_{j}$ we need not define the derivative of $R\left( \cdot \right) $ at
the points $z=0$ and $z=1$ because with a continuously distributed variable in $\mathbf{x}$ and non-zero coefficient it takes on these two values with probability zero. Therefore, the APE is  \vspace{-3mm}
\begin{equation*}
\alpha _{j}\equiv E\left[ PE_{j}\left( \mathbf{x}\right) \right] =\beta
_{j}P\left( 0\leq \mathbf{x\beta }\leq 1\right) =\beta _{j}P\left( 0<\mathbf{%
x\beta }\leq 1\right) =\beta _{j}\left[ F_{\mathbf{x\beta }}\left( 1\right)
-F_{\mathbf{x\beta }}\left( 0\right) \right], \tag{2.2} 
\label{alphaj}  \vspace{-3mm}
\end{equation*}%
where $F_{\mathbf{x\beta }}\left( \cdot \right) $ is the CDF of $\mathbf{%
x\beta }$.

There are some simple but useful observations about (\ref{alphaj}). First, $\alpha
_{j}$ always has the same sign as $\beta _{j}$. Second, because $F_{\mathbf{%
x\beta }}\left( 1\right) -F_{\mathbf{x\beta }}\left( 0\right) \leq 1$, $\left\vert \alpha _{j}\right\vert \leq \left\vert \beta _{j}\right\vert$; with wide support for $\mathbf{x\beta }$, $\alpha _{j}$ can be much smaller
in magnitude than $\beta _{j}$. Moreover, $\alpha _{j}=\beta _{j}$ if and
only if  \vspace{-3mm}
\begin{equation*}
P\left( \mathbf{x\beta }\in \left[ 0,1\right] \right) =1, \tag{2.3} \label{within01}  \vspace{-3mm}
\end{equation*}%
which means the support of $\mathbf{x\beta }$ is inside the unit interval.
This is essentially the condition used by Horrace and Oaxaca (2006) to
conclude that the OLS estimator in a linear regression is unbiased and consistent for $%
\mathbf{\beta }$. Our goal here is to compare the OLS estimators with the
APEs in the general case where $P\left( \mathbf{x\beta }\in \left[ 0,1\right]
\right) <1$; the H-O condition is then a special case where the index
coefficient, $\beta _{j}$, is identical to the APE, $\alpha _{j}$.

There is a third set of parameters important for the discussion, and those
are the linear projection parameters. Assume that the $x_{j}$ have finite
second moments and that the $K\times K$ matrix $E\left( \mathbf{x}^{\prime }%
\mathbf{x}\right) $ is nonsingular; this simply rules out perfect
collinearity in the population. Then we can always define the $K\times 1$
vector $\mathbf{\gamma }$ as \vspace{-3mm}
\begin{equation*}
\mathbf{\gamma }=\left[ E\left( \mathbf{x}^{\prime }\mathbf{x}\right) \right]
^{-1}E\left( \mathbf{x}^{\prime }y\right). \vspace{-3mm}
\end{equation*}%
We then write the linear projection of $y$ on $\left( 1,x_{2},...,x_{K}\right) $ as. \vspace{-3mm}
\begin{equation*}
L\left( y|\mathbf{x}\right) =L\left( y|1,x_{2},...,x_{K}\right) =\gamma
_{1}+\gamma _{2}x_{2}+\cdots +\gamma _{K}x_{K}=\mathbf{x\gamma } \vspace{-3mm}.
\end{equation*}%
In understanding the H-O findings, and their limitations, it is important to
know that $\alpha _{j}$, $\beta _{j}$, and $\gamma _{j}$ are all
well-defined parameters and, in general, they are all different. Defining $%
\mathbf{\beta }$ and $\mathbf{\alpha }$ requires an underlying model for the response probability whereas
defining $\mathbf{\gamma}$ does not.

As is well known, under random sampling the OLS estimator consistently
estimates the parameters of the linear projection; see, for example,
Wooldridge (2010, Chapter 4.2). In other words, if we run the OLS
regression underlying LPM estimation, \vspace{-3mm}
\begin{equation*}
y_{i}\text{ on }1\text{, }x_{i2}\text{, ..., }x_{iK}\text{, }i=1,...,N, \vspace{-3mm}
\end{equation*}%
and obtain the $\hat{\gamma}_{j}$, then $\hat{\gamma}_{j}\overset{p}{%
\rightarrow }\gamma _{j}$. Again, this result holds free of any kind of
underlying model.

H-O study the consistency of the $\hat{\gamma}_{j}$ when considered as
estimators of $\beta _{j}$---the coefficients in the index. In other words,
their asymptotic analysis is the same as comparing the linear projection
parameters $\gamma _{j}$ to the index parameters $\beta _{j}$. Our view is
that this does usually not make much sense---for the same reason we do not study
consistency of the OLS estimator for the index parameters in, say, probit or
logit. If one explicitly models the response probability as a nonlinear
function of $\mathbf{x\beta }$ then one must recognize that nonlinearity
when defining the parameters of interest. When interest is in the effects of
the explanatory variables on the response probability---which describes
almost all modern usages of the LPM---it only makes sense to compare the
linear projection parameters to the APEs. In other words, we should ask:
When is $\gamma _{j}$ \textquotedblleft close\textquotedblright\ to $\alpha
_{j}$? This is not the same as studying when $\gamma _{j}$ is
\textquotedblleft close\textquotedblright\ to $\beta _{j}$ (except in the
special case where (\ref{within01}) holds).

Under the H-O ramp model we can write\vspace{-3mm}
\begin{equation*}
E\left( y|\mathbf{x}\right) =p\left( \mathbf{x}\right) =1\left[ 0\leq 
\mathbf{x\beta }\leq 1\right] \mathbf{x\beta }+1\left[ \mathbf{x\beta }>1%
\right] \vspace{-3mm}.
\end{equation*}%
If (\ref{within01}) holds then, with probability one, \vspace{-3mm}
\begin{equation*}
E\left( y|\mathbf{x}\right) =\mathbf{x\beta }=L\left( y|\mathbf{x}\right) , \vspace{-3mm}
\end{equation*}%
in which case $\alpha _{j}=\beta _{j}$ and the OLS estimators, $\hat{\gamma}%
_{j}$ are consistent for $\beta _{j}$ (which is the APE of $x_{j}$). If for
a random sample of size $N$, $\mathbf{x}_{i}\mathbf{\beta }\in \left[ 0,1%
\right] $ for all $i$, then \vspace{-3mm}
\begin{equation*}
E\left( y_{i}|\mathbf{x}_{1},\mathbf{x}_{2},...,\mathbf{x}_{n}\right) =%
\mathbf{x}_{i}\mathbf{\beta }, \vspace{-3mm}
\end{equation*}%
and it follows that the OLS estimators are conditionally unbiased for the $%
\beta _{j}$ -- the conclusion reached in H-O.

If (\ref{within01}) fails then $\beta _{j}$ measure the partial effect when $0\leq 
\mathbf{x\beta }\leq 1$, but this restriction depends on the unknown vector $%
\mathbf{\beta }$. If $P\left( \mathbf{x\beta }\in \left[ 0,1\right] \right)
<1$ then the $\beta _{j}$ need not be very useful as summary measures of the
partial effects. In the next section we discuss when the LP\ parameters
identify APEs, with (\ref{within01}) being a special case.

In the next section, it is useful to observe that the response probability
in (\ref{HOspec}) can be derived from a latent variable formulation. Suppose that \vspace{-3mm}
\begin{align}
    y^{\ast } &=\mathbf{x\beta }+u  \tag{2.4} \label{latent1}, \\
u|\mathbf{x} &\sim \mathrm{Uniform}\left( 0,1\right) \tag{2.5} \label{latent2}, \\
y &=1\left[ y^{\ast }>0\right] \tag{2.6} \label{latent3} \vspace{-3mm}.
\end{align}
Because the CDF of $u$ is identical to the ramp function $R\left( \cdot
\right) $, it follows immediately that (\ref{latent1}), (\ref{latent2}), and (\ref{latent3}) lead to the response probability in (\ref{HOspec}).

\section{When are the Linear Projection Parameters \\ Identical to the APEs?}

In addition to being easy to interpret and readily extending to cases with
endogenous explanatory variables and unobserved heterogeneity, empirically
the OLS estimates of the LPM\ are often similar to the corresponding APEs
from nonlinear index models---particularly logit or probit. \citealp[Section 15.6]{wooldridge2010econometric} provides a discussion based on a results of Stoker
(1986) that helps one understand these empirical findings. Here we\ expand
that discussion to allow for an extension of the H-O framework.

It is useful to start with a general setting. Consider an index model\vspace{-3mm}
\begin{equation*}
P\left( y=1|\mathbf{x}\right) =G\left( \mathbf{x\beta }\right) =G\left(
\beta _{1}+\beta _{2}x_{2}+\cdots +\beta _{K}x_{K}\right)  \vspace{-3mm}
\end{equation*}%
where $G:\mathbb{R}\rightarrow \left[ 0,1\right] $. As argued in \citealp[Section 15.6]{wooldridge2010econometric}, the results of \cite{Stoker1986} imply that, if $\left(
x_{2},...,x_{K}\right) $ has a multivariate normal distribution and $G\left(
\cdot \right) $ is differentiable almost everywhere on $\mathbb{R}$ (with
respect to Lebesgue measure), then \vspace{-3mm}
\begin{equation*}
\gamma _{j}=\beta _{j}E\left[ g\left( \mathbf{x\beta }\right) \right] \equiv
\alpha _{j}\text{, }j=2,...,K, \vspace{-3mm}
\end{equation*}%
where $\gamma _{j}$ is the slope coefficients on $x_{j}$ in $L\left( y|%
\mathbf{x}\right) =\mathbf{x\gamma }$, $g\left( \cdot \right) $ is the
almost everywhere derivative of $G\left( \cdot \right) $, and $\alpha _{j}$
is the APE. The ramp function $R\left( \cdot \right) $ is differentiable
everywhere except at zero and one, and so it satisfies Stoker's (1986)
assumptions. The result is that OLS consistently estimates the APEs, $\alpha_{j}$, even though the $\alpha _{j}$ are attenuated versions of the $\beta_{j}$: \vspace{-3mm}
\begin{equation*}
\alpha _{j}=\beta _{j}\cdot P\left( 0\leq \mathbf{x\beta }\leq 1\right) 
\end{equation*}%
This equality holds even when $P\left( 0\leq \mathbf{x\beta }\leq 1\right) $
can be very close to zero. \cite{Horrace2006}, and many papers citing
their findings, focus on the inconsistency of OLS for $\beta _{j}$, failing
to recognize that the OLS estimators from the linear model could be
consistent for the more interesting quantities, the $\alpha _{j}$. This
point is key to our argument: If the model of the response probability is
nonlinear so that $0\leq p\left( \mathbf{x}\right) \leq 1$ is ensured, one
should study estimation of APEs, not underlying index parameters.

Clearly the assumption of multivariate normality of $\mathbf{x}$ is too
restrictive to be widely applicable. Nevertheless, the results of \cite{Stoker1986} are suggestive, especially when combined with \cite{Ruud1983}. Ruud studies smooth nonlinear function forms that never hit the endpoints of the unit interval, like probit and logit. In these cases, quasi-MLE identifies the index coefficients up to scale. If $\mathbf{x}$ has a centrally symmetric distribution---of which the multivariate normal is a special case---then Ruud's (1983) conditions hold. In the next section, we will find that when $\mathbf{x}$ are symmetrically distributed with small variance (not too spread out), the average partial effects are still approximated well by the LP parameters; as the variance increases, however, the approximation breaks down.

To facilitate further discussion, including the simulations in the next section, it is helpful to modify and extend the H-O setup. In particular, write \vspace{-3mm}
\begin{align*}
    y^{\ast } &=\mathbf{x\beta }+u \\
u|\mathbf{x} &\sim \mathrm{Uniform}\left( -a,a\right) \tag{3.1} \label{HOextend} \\
y &= 1\left[ y^{\ast }>0\right] \vspace{-3mm}
\end{align*}
for some $a>0$. Compared with H-O, we have shifted the intercept so that $u$ has a symmetric distribution about its mean of zero. Also, we allow $u$ to have narrow or wide support, depending on $a$. While the latent error support is not identified, it is a convenient device for generating data where the unit interval for probabilities is binding to varying degrees. The CDF for the $\mathrm{%
Uniform}\left( -\sqrt{3},\sqrt{3}\right) $ distribution, which has unit
variance, is graphed in Figure 2.

\begin{center}
    \includegraphics[scale=0.4]{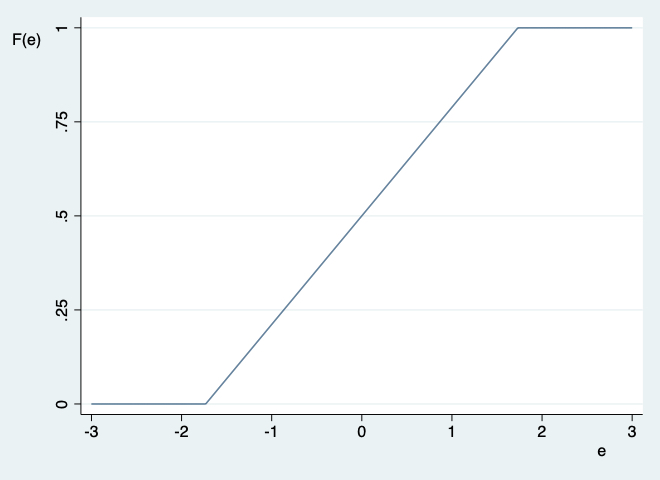} \\
    \textbf{Figure 2}: The CDF of $y$ with $u|x \sim \mathrm{%
U}\left( -\sqrt{3},\sqrt{3}\right) $.
\end{center}

Given the latent variable model in (\ref{HOextend}), we can derive the response
probability:  \vspace{-3mm}
\begin{eqnarray*}
p\left( \mathbf{x}\right) &\equiv &P\left( y=1|\mathbf{x}\right) =P\left(
y^{\ast }\geq 0|\mathbf{x}\right) =P\left( u\geq -\mathbf{x\beta }|\mathbf{x}%
\right) =P\left( u\leq \mathbf{x\beta }|\mathbf{x}\right) =F_{u}\left( 
\mathbf{x\beta }\right) \\
&=&0\text{ if }\mathbf{x\beta }<-a \\
&=&\frac{\mathbf{x\beta }+a}{2a}\text{ if }-a\leq \mathbf{x\beta }\leq a \\
&=&1\text{ if }\mathbf{x\beta }>a  \vspace{-3mm}
\end{eqnarray*}%
We write this function as $F_{u}\left( \mathbf{x\beta }\right) \equiv
R_{a}\left( \mathbf{x\beta }\right) $, which is a ramp function that is
nondifferentiable at $-a$ and $a$. For an $x_{j}$ with a positive
coefficient, the response probability has the same shape as in Figure 1. As $%
a$ increases relative to $\mathbf{\beta}$, the response probability is linear over more of the support of 
$\mathbf{x}$. If  \vspace{-3mm}
\begin{equation} 
P\left( -a\leq \mathbf{x\beta }\leq a\right) = 1  \tag{3.2} \label{withinaa}  \vspace{-3mm}
\end{equation}%
then, with probability one, $R_{a}\left( \mathbf{x\beta }\right) =\left( 
\mathbf{x\beta }+a\right) /2a$, a linear function of $\mathbf{x}$. In this
case, the partial effects are constant and equal to $\beta _{j}/2a$, $j=2,...,K$. These
are also the linear projection parameters $\gamma _{j}$ and so OLS
consistently estimates the APEs under (\ref{withinaa}).

If $x_{j}$ is a continuous variable, we are interested in the APE defined as
a derivative, which exists with probability one when $\mathbf{x\beta }$ is
continuous. At $\mathbf{x\beta }\in \left\{ -a,a\right\} $ the definition of
the partial effect is immaterial. To be concrete, take  \vspace{-3mm}
\begin{equation*}
PE_{j}\left( \mathbf{x}\right) =\frac{\beta _{j}}{2a}\cdot 1\left[ -a\leq 
\mathbf{x\beta }\leq a\right].  \vspace{-3mm}
\end{equation*}%
Notice that $PE_{j}\left( \mathbf{x}\right) =0$ if $\mathbf{x\beta }<-a$ or $%
\mathbf{x\beta }>a$ because we are on one of the flat parts of the ramp.
This feature of $PE_{j}\left( \mathbf{x}\right) $ is taken into account in
computing the APE:  \vspace{-3mm}
\begin{equation*}
\alpha _{j}=E\left[ PE_{j}\left( \mathbf{x}\right) \right] =\frac{\beta _{j}%
}{2a}\cdot P\left( -a\leq \mathbf{x\beta }\leq a\right)  \vspace{-3mm}
\end{equation*}%
Furthermore, the previous results based on Stoker (1986) still hold: If $%
\left( x_{2},...,x_{K}\right) $ is multivariate normal then, again letting $%
\gamma _{j}$ denote the LP\ parameter,  \vspace{-3mm}
\begin{equation*}
\gamma _{j}=\alpha _{j} \tag{3.3} \label{stoker}  \vspace{-3mm}
\end{equation*}%
Again, it is important to understand that (\ref{stoker}) holds even if $P\left(
-a\leq \mathbf{x\beta }\leq a\right) $ is close to zero (with zero being
ruled out); consequently, H-O's discussion about the amount of inconsistency
in the OLS estimators if $P\left( -a\leq \mathbf{x\beta }\leq a\right) <1$
is incomplete because they focus on $\beta _{j}$, not $\alpha _{j}$%
. In the extended model (\ref{HOextend}), depending on the values of $a$ and $P\left(
-a\leq \mathbf{x\beta }\leq a\right) $, $\left\vert \alpha _{j}\right\vert $
need not be smaller than $\left\vert \beta _{j}\right\vert $. The case that
aligns with H-O is $a=1/2$---so that the $Uniform\left( 0,1\right) $
distributed has just been shifted to have zero mean---in which case $%
\left\vert \alpha _{j}\right\vert \leq \left\vert \beta _{j}\right\vert $,
and the difference between $\alpha _{j}$ and $\beta _{j}$ can be large. It
is easily seen that $\left\vert \alpha _{j}\right\vert <\left\vert \beta
_{j}\right\vert $ for any $a\geq 1/2$.

We can also define APEs for discrete changes in the explanatory variables.
For example, if $x_{K}$ is binary, its partial effect is  \vspace{-3mm}
\begin{equation*}
PE_{K}\left( \mathbf{x}\right) =R_{a}\left( \beta _{1}+\beta
_{2}x_{2}+\cdots +\beta _{K-1}x_{K-1}+\beta _{K}\right) -R_{a}\left( \beta
_{1}+\beta _{2}x_{2}+\cdots +\beta _{K-1}x_{K-1}\right) ,  \vspace{-3mm}
\end{equation*}%
which corresponds to setting $x_{K}$ at its two values and obtaining the
difference in probabilities. Averaging across the joint distribution of the
other explanatory variables gives the APE:  \vspace{-3mm}
\begin{equation*}
\alpha _{K}\equiv APE_{K}=E\left[ PE_{K}\left( \mathbf{x}\right) \right]  \vspace{-3mm}
\end{equation*}%
If $x_{K}$ is a binary intervention or treatment indicator, $\alpha _{K}$ is
the average treatment effect.

Other than the case of multivariate normality of $\left(
x_{2},...,x_{K}\right) $, there is another case where the LP parameters, $%
\gamma _{j}$, $j=2,...,K$, equal the APEs: $x_{2}$, ..., $x_{K}$ are
mutually exclusive binary indicators that, along with a base group given by $%
x_{2}=x_{3}=\cdots =x_{K}=0$, are exhaustive. See \citealp[Section 3.1.4]{angrist2009mostly} and \citealp[Section 15.2]{wooldridge2010econometric}. If $x_{1}=1$ denotes the base group then
the APEs are simply  \vspace{-3mm}
\begin{equation*}
\alpha _{j}=E\left( y|x_{j}=1\right) -E\left( y|x_{1}=1\right) \text{, }%
j=2,...,K,  \vspace{-3mm}
\end{equation*}%
and these are identical to the corresponding LPM coefficients.

Beyond the extreme cases described here, there appears to be no general
theory to determine when the LP coefficients will be the same or
\textquotedblleft close\textquotedblright\ to the APEs. Many empirical
applications include a combination of continuous, discrete, and even mixed
explanatory variables. Rarely do these have marginal symmetric
distributions, let alone a symmetric joint distribution. Plus, such
explanatory variables often appear appear as quadratics, interactions, and
other functional forms---which also do not have symmetric distributions. In
Section 5, we use simulations to shed light on when the the LPM
coefficients closely approximate the APEs---and when they do not. First, however, we describe an estimator which is consistent under the ramp model.

\section{The NLS Estimator of the Ramp Model}

We have already seen how if $P(\mathbf{x}_i\mathbf{\beta} \in \left[0,1\right])=1$, then OLS is consistent for the $\beta_{j}$, which are equal to the APEs $\alpha_j$ in the case of a continuous covariate $x_j$ under model (\ref{HOspec}). If the probability that $\mathbf{x}_i\mathbf{\beta}$ lies outside the unit interval is nonzero, then OLS is no longer consistent for the $\beta_{j}$, and it may or may not approximate the $\alpha_j$ depending on the distribution of $\mathbf{x}$. In addition to probit and logit quasi-MLE, it makes sense to consider an estimator which is consistent if the ramp model is true. Of course, Bernoulli MLE using the ramp model as the conditional response probability is not feasible because the log-likelihood is not defined for $\mathbf{x\beta}\notin (0,1)$. Instead, we consider nonlinear least squares (NLS) using the piecewise ramp function $R(\mathbf{x}_i\mathbf{\beta})$ from (\ref{HOspec}) as the conditional mean. In addition, since there may not be much justification to think the ramp function is the true response probability, we allow for general misspecification. Therefore, we define $\mathbf{\beta}_o$ as the pseudo-true value in the sense that $\mathbf{\beta}_o$ is the unique solution to \vspace{-3mm}
\begin{align*}
    \underset{\mathbf{\beta}}{\text{min }} E\left[\left(y_i - R(\mathbf{x}_i\mathbf{\beta})\right)^2\right] \equiv  \underset{\mathbf{\beta}}{\text{min }} Q(\mathbf{\beta}). \tag{4.1} \label{pseudo_ramp} \vspace{-3mm}
\end{align*}
We say that the model is misspecified if there is no such $\mathbf{\beta}$ such that $E[y|\mathbf{x}] = R(\mathbf{x}\mathbf{\beta})$. By construction, $\mathbf{\beta}_o$ is the true coefficient when the model is correctly specified and otherwise we view $R(\mathbf{x}\mathbf{\beta}_o)$ as the best mean squared error approximation to $E[y|x] $ over all ramp functions $R(\mathbf{x}\mathbf{\beta})$.

As a sample analogue of (\ref{pseudo_ramp}), we define the objective function $Q_N (\mathbf{\beta})$ as  \vspace{-3mm}
\begin{align*} 
    Q_N (\mathbf{\beta}) \equiv& \frac{1}{N} \sum_{i=1}^N \left(y_i - R(\mathbf{x}_i\mathbf{\beta})\right)^2 \nonumber \\
    =& \frac{1}{N} \sum_{i=1}^N \left(y_i^21\left\{\mathbf{x}_i\mathbf{\beta}\leq 0\right\} + \left(y_i - \mathbf{x}_i\mathbf{\beta}\right)^2 1\left\{\mathbf{x}_i\mathbf{\beta}\in (0,1)\right\} + (y_i -1)^2 1\left\{\mathbf{x}_i\mathbf{\beta}\geq 1\right\}\right), \vspace{-3mm}
\end{align*}
where $N$ is the sample size. We define the NLS estimator $\hat{\mathbf{\beta}}$ as  \vspace{-3mm}
\begin{equation*}
    \hat{\mathbf{\beta}}\equiv \text{arg}\underset{\mathbf{\beta}}{\text{min }} \, Q_N (\mathbf{\beta}).  \vspace{-3mm}
\end{equation*}

The following theorem gives the consistency of the NLS estimator for the pseudo-true value, allowing for misspecification of the conditional mean model.
\begin{theorem}
    Let $\left\{y_i, \mathbf{x}_i\right\}_{i=1}^{\infty}$ be an i.i.d. sequence with $y$ only taking on values zero and one, and let $R: \mathbb{R} \to [0,1]$ be the ramp function defined in (\ref{HOspec}). Suppose $\mathbf{\beta}\in\mathbf{\mathcal{B}}$ such that $\mathbf{\mathcal{B}}\subset \mathbb{R}^K$ is compact, and $\mathbf{\beta}_o$ is identified in the sense that $\forall \, \mathbf{\beta} \in \mathbf{\mathcal{B}}, \mathbf{\beta}\neq \mathbf{\beta}_o$,
    \[
    E\left[\left(y_i - R(\mathbf{x}_i\mathbf{\beta}_o)\right)^2\right]
    < E\left[\left(y_i - R(\mathbf{x}_i\mathbf{\beta})\right)^2\right]
    \]
    Then, $\hat{\mathbf{\beta}} \overset{p}{\rightarrow} \mathbf{\beta}_o$ as $N\to \infty$.
\end{theorem}
The consistency result of Theorem 1 follows directly from Theorem 12.2 of \cite{wooldridge2010econometric}.

If $\mathbf{x}$ contains a continuously distributed $x_j$ and $\beta_{jo}$ is nonzero, then the probability of $\mathbf{x}_i\mathbf{\beta}_o$ being equal to 0 or 1 is zero. Then, with suitable moment conditions on $\mathbf{x}$ (so the Leibniz integral rule applies), the FOC of the (\ref{pseudo_ramp}) is well defined with probability 1 as follows: \vspace{-3mm}
\begin{align*}
    E\left[\mathbf{x}_i'u_i 1\{\mathbf{x}_i\mathbf{\beta}_o\in(0,1)\}\right] = 0 ,\tag{4.2} \label{score} \vspace{-3mm}
\end{align*}
where $u_i(\mathbf{\beta}) = y_i - R(\mathbf{x}_i\mathbf{\beta})$ and $u_i\equiv u_i(\mathbf{\beta}_0)$. Define the score function for random draw $i$:  \vspace{-3mm}
\begin{align*}
    \mathbf{s}_i(\mathbf{\beta}) = -\mathbf{x}_i'u_i(\mathbf{\beta}) 1\{\mathbf{x}_i\mathbf{\beta}\in (0,1)\}. \vspace{-3mm}
\end{align*}
Then, $\mathbf{\beta}_o$ solves $E[\mathbf{s}_i(\mathbf{\beta}_o)] = 0$. The variance-covariance matrix of $\mathbf{s}_i(\mathbf{\beta})$ is  \vspace{-3mm}
\begin{align*}
     \mathbf{\Omega}(\mathbf{\beta})=E\left[\mathbf{x}_i'\mathbf{x}_iu_i(\mathbf{\beta})^2 1\left\{\mathbf{x}_i\mathbf{\beta}\in(0,1)\right\} \right]. \tag{4.3} \label{omega} \vspace{-3mm}
\end{align*}
The natural definition of the Jacobian of $\mathbf{s}_i(\mathbf{\beta})$ is \vspace{-3mm}
\begin{align*}
     \mathbf{A}_i(\mathbf{\beta}) = \mathbf{x}_i'\mathbf{x}_i 1\left\{\mathbf{x}_i\mathbf{\beta}\in(0,1)\right\}.  \vspace{-3mm}
\end{align*}
For the similar reason as (\ref{score}), the Hessian of $Q(\mathbf{\beta})$ is well-defined with probability 1 at $\mathbf{\beta}_o$ as follows  \vspace{-3mm}
\begin{align*}
     \mathbf{A}(\mathbf{\beta}_o) = E\left[\mathbf{x}_i'\mathbf{x}_i 1\left\{\mathbf{x}_i\mathbf{\beta}_o\in(0,1)\right\} \right]. \tag{4.4} \label{Hessian} \vspace{-3mm}
\end{align*}

Note that (\ref{omega}) and (\ref{Hessian}) are the same whether the conditional mean model is correctly specified or not. Therefore, the following asymptotic distribution result allows for misspecification of the model.

\begin{theorem}
    Suppose that the assumptions from Theorem 1 hold, and (i) $\mathbf{\beta}_o$ is an interior point of $\mathbf{\mathcal{B}}$; (ii) $\mathbf{x}_i$ contains a continuously distributed random variable with a nonzero coefficient; (iii) $E\left\Vert \mathbf{x}_i\right\Vert^2<\infty$ and $E\left[\mathbf{x}_i'\mathbf{x}_i 1\left\{\mathbf{x}_i\mathbf{\beta}_o\in(0,1)\right\} \right]>0$, where $\Vert.\Vert$ denotes the $l^2-norm$. Then, as $N\to\infty$,  \vspace{-3mm}
    \begin{align*}
       \sqrt{N}\left(\hat{\mathbf{\beta}}-\mathbf{\beta}_o\right)\Rightarrow \mathbb{N}(0,\mathbf{A}(\mathbf{\beta}_o)^{-1} \mathbf{\Omega}(\mathbf{\beta}_o) \mathbf{A}(\mathbf{\beta}_o)^{-1}) . \vspace{-3mm}
    \end{align*}
\end{theorem}

The proof of Theorem 2 is given in the Appendix. The asymptotic normality results does not follow directly from the M-estimator due to the non-smoothness of the objective function. We therefore leverage an asymptotic normality result for estimators with non-smooth objective function from \cite{Newey1994}.

To estimate $\mathbf{\beta}_o$, H-O suggest running OLS on a trimmed sample (i.e., those observations for which initial OLS fitted values are inside the unit interval) to reduce bias. We find in practice that a single round of trimming may not reduce the bias for the APEs in the cases where OLS is not consistent for them. However, we find an iterative trimming OLS procedure (ITO) does reduce the bias for estimating APEs, as well as $\mathbf{\beta}_o$.\footnote{The procedure goes: 1) estimate the LPM by OLS. 2) Compute fitted values. 3) Drop observations with fitted values outside the unit interval, and 4) Repeat starting at 1) until no further observations are dropped.} In fact, we find in simulations that the NLS estimates are numerically the same as the ITO estimates up to machine precision.\footnote{With some DGP, it was occasionally necessary to specify OLS starting values for the NLS function evaluator for this result to hold.} It turns out that ITO is implicitly minimizing the NLS sample objective function using the OLS estimates as starting values and following the Newton-Raphson numerical method, which is iterative (see \citealp[Section 12.7.1]{wooldridge2010econometric}). Given an estimate $\mathbf{\beta}^{\left\{g\right\}}$, the next iteration is given (using our notation) by   \vspace{-3mm}
\begin{align*}\mathbf{\beta}^{\left\{g+1\right\}} &= \mathbf{\beta}^{\left\{g\right\}} - \left[N^{-1}\sum_{i=1}^N \mathbf{A}_i(\mathbf{\beta}^{\left\{g\right\}}) \right]^{-1} N^{-1} \sum_{i=1}^N \mathbf{s}_i (\mathbf{\beta}^{\left\{g\right\}}) \\
&=\mathbf{\beta}^{\left\{g\right\}} + \left[N^{-1}\sum_{i=1}^N \mathbf{x}_i'\mathbf{x}_i 1\left\{\mathbf{x}_i\mathbf{\beta}^{\left\{g\right\}}\in(0,1)\right\} \right]^{-1} N^{-1} \sum_{i=1}^N \mathbf{x}_i' \left(y_i - \mathbf{x}_i\mathbf{\beta}^{\left\{g\right\}}\right)1\left\{\mathbf{x}_i\mathbf{\beta}^{\left\{g\right\}}\in(0,1)\right\} \\
&=\left[N^{-1}\sum_{i=1}^N \mathbf{x}_i'\mathbf{x}_i 1\left\{\mathbf{x}_i\mathbf{\beta}^{\left\{g\right\}}\in(0,1)\right\} \right]^{-1} N^{-1} \sum_{i=1}^N \mathbf{x}_i' y_i 1\left\{\mathbf{x}_i\mathbf{\beta}^{\left\{g\right\}}\in(0,1)\right\} .  \vspace{-3mm}
\end{align*}
The second equality above substitutes our expressions for $\mathbf{s}_i (\mathbf{\beta})$ and $\mathbf{A}_i(\mathbf{\beta})$ and uses the fact that $R(\mathbf{x}_i\mathbf{\beta})=\mathbf{x}_i\mathbf{\beta}$ for $\mathbf{x}_i\mathbf{\beta}\in (0,1)$. This shows that $\mathbf{\beta}^{\left\{g+1\right\}}$ is simply the OLS estimator on the sample with $\mathbf{x}_i\mathbf{\beta}^{\left\{g\right\}}\in(0,1)$. 

As a consequence, the preceding consistency and asymptotic normality results for the NLS estimator justify using the ITO procedure to reduce the OLS bias. However, it is worth mentioning that, at least in Stata, the pre-loaded NLS solver (the``nl'' command) may have a performance advantage over ITO in practice. We find in simulations that ITO can result in a dead loop when only a very small portion of observations are left for estimation after iterative trimming. The pre-loaded NLS algorithm continues to work well in those cases.

Taking the sample analogue of the asymptotic variance from Theorem 2, we define a variance estimator of $\sqrt{N}(\hat{\mathbf{\beta}}-\mathbf{\beta}_o)$ as  \vspace{-3mm}
\begin{align*}
    \mathbf{\hat{V}} = \mathbf{A}_N(\hat{\mathbf{\beta}})^{-1} \mathbf{\Omega}_N(\hat{\mathbf{\beta}}) \mathbf{A}_N(\hat{\mathbf{\beta}})^{-1},  \vspace{-3mm}
  \end{align*}  
where $\mathbf{A}_N(\hat{\mathbf{\beta}})=N^{-1}\sum_{i=1}^N \mathbf{x}_i'\mathbf{x}_i 1\{\mathbf{x}_i\hat{\mathbf{\beta}}\in(0,1)\}$, $\mathbf{\Omega}_N(\hat{\mathbf{\beta}}) = N^{-1}\sum_{i=1}^N \mathbf{x}_i'\mathbf{x}_i \hat{u}^2_i 1\{\mathbf{x}_i\mathbf{\hat{\beta}}\in(0,1)\}$, and $\hat{u}_i = y_i - R(\mathbf{x}_i\hat{\mathbf{\beta}})$. Standard errors are obtained the usual way from $\mathbf{\hat{V}}/N$. The next theorem gives the consistency result of the variance estimator. 
    
\begin{theorem}
    Under the same assumption of Theorem 2 and $E\Vert x\Vert^4<\infty$, as $N\to\infty$, $\mathbf{\hat{V}} \stackrel{p}{\to} \mathbf{A}(\mathbf{\beta}_o)^{-1} \mathbf{\Omega}(\mathbf{\beta}_o)\mathbf{A}(\mathbf{\beta}_o)^{-1} $.
\end{theorem}

The proof of Theorem 3 is given in the Appendix. As before, we are interested in the APE. Consider the best ramp approximation in (\ref{pseudo_ramp}), the APE of a continuous random variable $x_k$ is defined as  \vspace{-3mm}
\begin{align*}
    APE_k = E\left[\frac{\partial R(\mathbf{x}_i\mathbf{\beta}_o)}{\partial x_k}\right] = \beta_{ko} P\left(\mathbf{x}_i\mathbf{\beta}_o\in(0,1)\right). \vspace{-3mm}
\end{align*}
A sample-analogue estimator of the APE is then given by  \vspace{-3mm}
\begin{align*}
    A\widehat{P}E_k = \hat{\mathbb{\beta}}_k \frac{1}{N} \sum_{i=1}^N 1\left\{\mathbf{x}_i \hat{\mathbf{\beta}}\in(0,1)\right\} \vspace{-3mm}
\end{align*}
Define $g(\mathbf{x}_i,\mathbf{\beta}) = \beta_{k}1\{\mathbf{x}_i,\mathbf{\beta}_o\}$, $\delta_o = E[g(\mathbf{x}_i,\mathbf{\beta}_o)]$, and $\mathbf{G}_o=\nabla_{\mathbf{\beta}}g(\mathbf{x}_i,\mathbf{\beta}_o)$. Following problem 12.17 of Wooldridge (2010), the asymptotic variance of the estimated APE is given by \vspace{-3mm}
\begin{align*}
    AVar\left(\sqrt{N}\left(A\widehat{P}E_k - APE_k\right)\right) = Var\left(g(\mathbf{x}_i,\mathbf{\beta}_o)-\delta_o-\mathbf{G}_o \mathbf{A}(\mathbf{\beta}_o)^{-1} \mathbf{s}_i(\mathbf{\beta}_o)\right), \vspace{-3mm}
\end{align*}
where $\mathbf{G}_o$ is a $1\times K$ vector with the $k^{th}$ element being $p_o\equiv P\left(\mathbf{x}_i\mathbf{\beta}_o\in(0,1)\right)$ and all else $0$. The asymptotic variance can be estimated by the sample variance of $g(\mathbf{x}_i,\hat{\mathbf{\beta}})-\hat{\delta}-\widehat{\mathbf{G}}\mathbf{A}_N(\hat{\mathbf{\beta}})^{-1} \mathbf{s}_i (\hat{\mathbf{\beta}}) $, where $\hat{\delta} = \frac{1}{N}\sum_{i=1}^N g(\mathbf{x}_i,\hat{\mathbf{\beta}})$, $\widehat{\mathbf{G}}$ is a $1\times K$ vector with the $k^{th}$ element being $\hat{p} = \frac{1}{N}\sum_{i=1}^N 1\left\{\mathbf{x}_i\hat{\mathbf{\beta}}\in(0,1)\right\}$.

The APE for a discrete random variable $x_k$ can be defined as \vspace{-3mm}
\begin{align*}
    APE_k = E\left[R(\mathbf{x}_{i,-k}\mathbf{\beta}_{-ko}+\beta_{ko}) - R(\mathbf{x}_{i,-k}\mathbf{\beta}_{-ko})\right]. \vspace{-3mm}
\end{align*}
A sample analogue estimator of $APE_k$ is given by \vspace{-3mm}
\begin{align*}
    A\widehat{P}E_k = \frac{1}{N} \sum_{i=1}^N R(\mathbf{x}_{i,-k}\hat{\mathbf{\beta}}_{-k}+\hat{\beta}_{k}) - R(\mathbf{x}_{i,-k}\hat{\mathbf{\beta}}_{-k}). \vspace{-3mm}
\end{align*}
The asymptotic variance can be found and estimated in a similar manner as the continuous case.

\section{Simulations}
In this section we present several Monte Carlo simulations that provide
insights into the behavior of different modeling/estimation approaches. 
The LPM is estimated by OLS and the ramp function is estimated by NLS. For NLS, the average partial effects are estimated based on averages of derivatives and differences of the ramp function. These resemble the familiar formulas for the linear model, though the individual unit partial effects need to be scaled by $1\left[0\leq \widehat{y} \leq 1\right]$ before averaging, where $\widehat{y}$ corresponds to the fitted values for each estimator. The logit and probit parameters are estimated by the (quasi-) maximum likelihood estimator, and then the average partial effects are estimated using the usual APE formulas. We further consider a nonparametric model estimated by the local linear estimator and the APEs are estimated by the sample average of the partial effects. We used Stata\textsuperscript{\textregistered}17 for simulation. The Stata code is available upon request. 

Initially, the true models take the form (we are dropping $o$ on beta here) \vspace{-3mm}
\begin{equation*}
y=1\left[ \beta _{0}+\beta _{1}x_{1}+\beta _{2}x_{2}+u>0\right] , \tag{5.1} \label{sim_y_noint} \vspace{-3mm}
\end{equation*}%
where $u$ is independent of $\left( x_{1},x_{2}\right) $ with $u\sim
Uniform\left( -a,a\right) $ for $a>0$. The choice of $a$ is important
because it governs how close to linear is the response probability for a given $\mathbf{\beta}$. The variable $x_{1}$ is continuous and $x_{2}$ is binary; they are
generated to be correlated. We indicate the intercept in the index function by 
$\beta _{0}$.

When $u\sim Uniform(-a, a)$, the ramp model is correctly specified, but the LPM is misspecified to varying degrees. For small $a$, the kinks in the ramp
function are binding and the LPM can provide a poor approximation to the
response probability. Naturally, the logit and probit models are always
misspecified in this case. As stated before, here we focus on the APEs rather than the underlying parameters or how well the models approximate the true response probability.

The sample size is $N=1,000$ and $1,000$ replications are used. The population (or true) APEs are not available in closed form, and so we
simulate these along with the estimators. In the tables to follow, the columns labeled ``Simulated Truth'' include the empirical means and standard deviations of the sample average partial effects at the true parameter values. We also simulate the probabilities  \vspace{-3mm}
\begin{equation*}
P\left( -a\leq \mathbf{x\beta }\leq a\right)  \vspace{-3mm}
\end{equation*}%
and  \vspace{-3mm}
\begin{equation*}
P\left( 0\leq \widehat{y} \leq 1\right),  \vspace{-3mm}
\end{equation*}%
where $\hat{y}$ refers to predicted values for the linear index. The first of these tells us how binding are the ramp function inflection points. The
second is practically relevant because researchers often check the fraction
of fitted values outside the unit interval as a way to determine the
adequacy of the LPM. The simulations show that having a large fraction of
fitted values in $\left[ 0,1\right] $ is neither necessary nor sufficient
for OLS to produce accurate estimates of the APEs.\footnote{H-O do note that consistency of OLS can no longer be shown as soon as one observation has a true index outside the unit interval.}

We also consider the case where an interaction term, $x_{1}\cdot x_{2}$, is
included in the model, and the researcher includes in interaction in the specification. In the Appendix, we choose $u\sim \mathrm{Normal}\left( 0,1\right) $ to compare the LPM with logit and probit when the probit model is correct.

\subsection{Symmetrically Distributed Explanatory Variables}

In the first design, $\left( x_{1},x_{2}\right) $ are generated as  \vspace{-3mm}
\begin{align*}
x_{1}&=v/\sqrt{2}+e/\sqrt{2}  \\
x_{2}&=1\left[ v/2+r>0\right] ,  \vspace{-3mm}
\end{align*}%
where $v$, $e$, and $r$ are independent standard normals. The index
parameters are set as  \vspace{-3mm}
\begin{equation*}
\left( \beta _{0},\beta _{1},\beta _{2}\right) =\left( 0.1,0.2,-0.3\right)  \vspace{-3mm}
\end{equation*}%
The binary outcome $y$ is generated as in (\ref{sim_y_noint}) with $u\sim \mathrm{Uniform}%
\left( -a,a\right) $, where $a\in \left\{ 1/4,1/2,1\right\} $. The case $%
a=1/2$ essentially corresponds to H-O. When $a=1$, the response probability
is essentially linear for these index parameter values. When $a=1/4$, the kinks are binding and $\mathbf{%
x\beta }$ is often outside the interval $\left[ -a,a\right] $.

Table 1 reports the findings when $a=1/2$. There is a small probability that
$\mathbf{x\beta }\notin \left[ -a,a\right] $ -- roughly, about 0.013.
Moreover, across all simulations, about $1.3\%$ of the OLS fitted values
are outside the unit interval. The pattern is clear: All the estimators of the APEs show very little bias and have the
same precision. This is true for the continuous variable, $x_{1}$, and the
binary variable, $x_{2}$. Note that this is not predicted by application of
the Stoker results because $x_{2}$ is a discrete variable. Nevertheless,
this table illustrates what is often observed in practice: the LPM
coefficients estimated by OLS are often close to the probit and logit APEs.

\begin{table}[th]
    \centering
    \small
    \begin{threeparttable}
    \begin{tabular}
[c]{cc|cccccc}%
\multicolumn{8}{c}{Table 1. No interaction, $x_1$ normal, $x_2$ sym. binary, $u\sim U(-0.5,0.5)$}\\ \hline\hline
\multicolumn{2}{c|}{$N=1000$} & Simulated & LPM & Ramp & Probit & Logit & Nonpar. \\
 &  & Truth*  & (OLS) & (NLS) & (QMLE) & (QMLE) & (LL) \\ \hline
\multirow{2}{4em}{$APE_1$} &mean  & 0.1975 & 0.1971 & 0.1972 &  0.1981 &  0.1969 &  0.1980 \\
                           &sd    & 0.0007 & 0.0132 & 0.0134 &  0.0131 &  0.0132 &  0.0141 \\ \hline
\multirow{2}{4em}{$APE_2$} &mean  & -0.2925 & -0.2960 & -0.2919& -0.2877 & -0.2851 & -0.2950 \\
                           &sd    & 0.0010 & 0.0299   & 0.0291 &  0.0283 &  0.0283 &  0.0291 \\ \hline
\multicolumn{2}{c|}{$P(y=1)$}     & 0.4509 &  & & &  &  \\
\multicolumn{2}{c|}{$P(0 \leq\widehat{y}\leq 1)$} &  & 0.9875  & 0.9860 &  1.0000 & 1.0000  & 0.9867\\ 
\multicolumn{2}{c|}{$P(-a\leq\mathbf{x\beta}\leq a)$} & 0.9873 &  &  &  &  &  \\ \hline 
\end{tabular}
\begin{tablenotes}
	\small
	\item *This column contains the empirical means and standard deviations of the sample average partial effects at the true parameter values.
	\end{tablenotes}
    \end{threeparttable}
\end{table}

The story does not change when the flat parts of the ramp function are
strongly binding. In Table 2, ${\small P}\left( -a\leq \mathbf{x\beta }\leq
a\right) $ is only about $0.78$, and about 11.8 percent of the OLS fitted
values are outside $\left[ 0,1\right] $. And yet, for estimating the APEs,
the LPM does essentially as well as probit and logit, with logit having
perhaps a bit less bias. But, given the
simulation error, these are not to be dwelt upon. Table 3 shows the case
where ${\small P}\left( -a\leq \mathbf{x\beta }\leq a\right) $ is exactly
one. We would expect the LPM to work very
well in this case, and it does---especially for the continuous variable $%
x_{1}$. What is, perhaps, more surprising is that probit and logit work just
as well, even though the true response probability is linear over the
support of $\mathbf{x\beta }$. These findings are a good reminder of why
statements such as \textquotedblleft the linear probability model is
preferred to probit because the latter assumes normality\textquotedblright\
are not just misleading: they are wrong. In the end, what we care about is
how well each approach approximates the partial effects on $P\left( y=1|%
\mathbf{x}\right) $. When we consider the APEs, all methods do well
even when the response probability has the peculiar ramp shape.

\begin{table}[th]
    \centering
    \small
    \begin{threeparttable}
    \begin{tabular}
[c]{cc|cccccc}%
\multicolumn{8}{c}{Table 2. No interaction, $x_1$ normal, $x_2$ sym. binary, $u\sim U(-0.25,0.25)$}\\ \hline\hline
\multicolumn{2}{c|}{$N=1000$} & Simulated & LPM & Ramp & Probit & Logit & Nonpar. \\
 &  & Truth*  & (OLS) & (NLS) & (QMLE) & (QMLE) & (LL) \\ \hline
\multirow{2}{4em}{$APE_1$} &mean  & 0.3124  &0.3137 & 0.3129 & 0.3152 & 0.3127 & 0.3270\\
                           &sd    & 0.0051  &0.0097 & 0.0098 & 0.0092 & 0.0096 & 0.0112\\ \hline
\multirow{2}{4em}{$APE_2$} &mean  & -0.4414 &-0.4774& -0.4418& -0.4424& -0.4385& -0.4567 \\
                           &sd    & 0.0058  &0.0229 & 0.0206 & 0.0198 & 0.0202 & 0.0240\\ \hline
\multicolumn{2}{c|}{$P(y=1)$}     & 0.4215 &   & & &  &  \\
\multicolumn{2}{c|}{$P(0 \leq\widehat{y}\leq 1)$} &  & 0.8822  & 0.7768 &  1.0000 & 1.0000  & 0.8770\\ 
\multicolumn{2}{c|}{$P(-a\leq\mathbf{x\beta}\leq a)$} & 0.7809 &   &  &  &  &  \\ \hline 
\end{tabular}
\begin{tablenotes}
	\small
	\item *This column contains the empirical means and standard deviations of the sample average partial effects at the true parameter values.
	\end{tablenotes}
    \end{threeparttable}
\end{table}

\begin{table}[th]
    \centering
    \small
    \begin{threeparttable}
    \begin{tabular}
[c]{cc|cccccc}%
\multicolumn{8}{c}{Table 3. No interaction, $x_1$ normal, $x_2$ sym. binary, $u\sim U(-1,1)$}\\ \hline\hline
\multicolumn{2}{c|}{$N=1000$} & Simulated & LPM & Ramp & Probit & Logit & Nonpar. \\
 &  & Truth* & (OLS) & (NLS) & (QMLE) & (QMLE) & (LL) \\ \hline
\multirow{2}{4em}{$APE_1$} &mean & 0.1000 &0.1001  & 0.1001 & 0.1001 & 0.1001 & 0.1002 \\
                           &sd   & 0.0000 &0.0159  & 0.0159 & 0.0159 & 0.0159 & 0.0163 \\ \hline
\multirow{2}{4em}{$APE_2$} &mean & -0.1500&-0.1509 & -0.1509& -0.1496& -0.1492& -0.1507 \\
                           &sd   & 0.0000 &0.0331  & 0.0331 & 0.0326 & 0.0325 & 0.0335 \\ \hline
\multicolumn{2}{c|}{$P(y=1)$}    & 0.4755  &      	 & 		  & 	    &         &         \\
\multicolumn{2}{c|}{$P(0 \leq\widehat{y}\leq 1)$}&&  1.0000 & 1.0000 & 1.0000	& 1.0000  & 0.9980  \\ 
\multicolumn{2}{c|}{$P(-a\leq\mathbf{x\beta}\leq a)$} & 1.0000  &  &  &  &  &  \\\hline  
\end{tabular}
\begin{tablenotes}
	\small
	\item *This column contains the empirical means and standard deviations of the sample average partial effects at the true parameter values.
	\end{tablenotes}
    \end{threeparttable}
\end{table}

We also generated the outcome $y$ using an interaction between $x_{1}$ and $%
x_{2}$, with $u$ still having a uniform distribution. Specifically,  \vspace{-3mm}
\begin{equation*}
y=1\left[ \beta _{0}+\beta _{1}x_{1}+\beta _{2}x_{2}+\beta _{3}\left(
x_{1}\cdot x_{2}\right) +u>0\right]  \vspace{-3mm}
\end{equation*}%

Remember, both $x_{1}$ and $x_{2}$ have symmetric distributions, but this
functional form falls outside Stoker's results because $x_{2}$ is discrete
and so is $x_{1}\cdot x_{2}$: it has a mass point at zero and is otherwise
continuous. Across a few parameter settings, the three approaches---where
the interaction term is included in the estimation---delivered similar estimated APEs
that were close to the sample ``true'' APEs. (As previously, probit, logit, and OLS approaches
use a misspecified response probability.) The parameters in that case are
set at  \vspace{-3mm}
\begin{equation*}
\left( \beta _{0},\beta _{1},\beta _{2},\beta _{3}\right) =\left(
0.1,0.2,-0.3,-0.3\right)  \vspace{-3mm}
\end{equation*}%

\begin{table}[thb!]
    \centering
    \small
    \begin{threeparttable}
    \begin{tabular}
[c]{cc|cccccc}%
\multicolumn{8}{c}{Table 4. With interaction, $x_1$ normal, $x_2$ sym. binary, $u\sim U(-0.5,0.5)$}\\ \hline\hline
\multicolumn{2}{c|}{$N=1000$} & Simulated & LPM & Ramp & Probit & Logit & Nonpar. \\
 &  & Truth* & (OLS) & (NLS) & (QMLE) & (QMLE) & (LL) \\ \hline
\multirow{2}{4em}{$APE_1$} &mean  & 0.0495 & 0.0493 & 0.0494 & 0.0492 & 0.0491 & 0.0495\\
                           &sd    & 0.0049 & 0.0143 & 0.0146 & 0.0144 & 0.0144 & 0.0156\\ \hline
\multirow{2}{4em}{$APE_2$} &mean  & -0.2989& -0.3002& -0.2990& -0.2941& -0.2929& -0.2997 \\
                           &sd    & 0.0099 & 0.0329 & 0.0326 & 0.0321 & 0.0322 & 0.0330\\ \hline
\multicolumn{2}{c|}{$P(y=1)$}     & 0.3898 &   &  &&  &  \\
\multicolumn{2}{c|}{$P(0 \leq\widehat{y}\leq 1)$} &  & 0.9938 & 0.9930 &  1.0000 & 1.0000  & 0.9946\\ 
\multicolumn{2}{c|}{$P(-a\leq\mathbf{x\beta}\leq a)$} & 0.9954 &  &  &  &  &  \\ \hline 
\end{tabular}
\begin{tablenotes}
	\small
	\item *This column contains the empirical means and standard deviations of the sample average partial effects at the true parameter values.
	\end{tablenotes}
    \end{threeparttable}
\end{table}

Tables 4 and 5 show the simulation findings. In Table 4, when the
support of $u$ is moderately wide, all methods perform about equally well.
They have little bias and their precisions are practically identical. In Table 5, when the support of $u$ narrows to $\left(
-1/4,1/4\right) $, the OLS estimation of LPM still does as well as the other estimations. These findings would seem
to go against conventional wisdom because there is a non-trivial fraction of fitted values outside the unit interval, about $0.15$. Finally, we note that the OLS estimator for $P(0\leq \widehat{y}\leq 1)$ can be severley biased for $P(-a \leq \mathbf{x\beta}\leq a)$.
\begin{table}[thb!]
    \centering
    \small
    \begin{threeparttable}
    \begin{tabular}
[c]{cc|cccccc}%
\multicolumn{8}{c}{Table 5. With interaction, $x_1$ normal, $x_2$ sym. binary, $u\sim U(-0.25,0.25)$}\\ \hline\hline
\multicolumn{2}{c|}{$N=1000$} & Simulated & LPM & Ramp & Probit & Logit & Nonpar. \\
 &  & Truth* & (OLS) & (NLS) & (QMLE) & (QMLE) & (LL) \\ \hline
\multirow{2}{4em}{$APE_1$} &mean  & 0.1104 & 0.1115 & 0.1108 & 0.1105 & 0.1102 & 0.1187\\
                           &sd    & 0.0079 & 0.0125 & 0.0127 & 0.0126 & 0.0128 & 0.0138\\ \hline
\multirow{2}{4em}{$APE_2$} &mean  & -0.5154& -0.5401& -0.5150& -0.5074& -0.5050& -0.5256 \\
                           &sd    & 0.0150 & 0.0275 & 0.0272 & 0.0268 & 0.0271 & 0.0280\\ \hline
\multicolumn{2}{c|}{$P(y=1)$}     & 0.3094 & & & & &  \\
\multicolumn{2}{c|}{$P(0 \leq\widehat{y}\leq 1)$} &  & 0.8503 & 0.6805 &  1.0000 & 1.0000  & 0.9640\\ 
\multicolumn{2}{c|}{$P(-a\leq\mathbf{x\beta}\leq a)$} & 0.6850 & &  &  &  &  \\ \hline 
\end{tabular}
\begin{tablenotes}
	\small
	\item *This column contains the empirical means and standard deviations of the sample average partial effects at the true parameter values.
	\end{tablenotes}
    \end{threeparttable}
\end{table}  \vspace{-3mm}

\subsection{Asymmetrically Distributed Explanatory Variables}
The story changes markedly when the distributions of $x_{1}$ and $x_{2}$ are
asymmetric. With $v$, $e$, and $r$ generated as before, $x_{1}$ and $x_{2}$
are now generated as \vspace{-3mm}
\begin{eqnarray*}
x_{1} &=&\exp \left( 0.5+v/2+e/2\right) \\
x_{2} &=&1\left[ -0.5+v/2+r>0\right] ,  \vspace{-3mm}
\end{eqnarray*}%
so that $x_{1}$ has a lognormal distribution. The variable $x_{2}$ is still
binary but the response probability is well below $0.5$. Tables 6 and 7
repeat the same experiments as Table 2 and Table 3, with $u\sim U(-0.25,0.25)$ and $u\sim U(-1,1)$ respectively, but with the covariates generated as above. The results for $u\sim U(-0.5,0.5)$ possess a similar pattern and thus omitted for brevity. The parameter values are, again,  \vspace{-3mm}
\begin{equation*}
\left( \beta _{0},\beta _{1},\beta _{2}\right) =\left( 0.1,0.2,-0.3\right)  \vspace{-3mm}
\end{equation*}

\begin{table}[th]
    \centering
    \small
    \begin{threeparttable}
    \begin{tabular}
[c]{cc|cccccc}%
\multicolumn{8}{c}{Table 6. No interaction, $x_1$ asym., $x_2$ asym. binary, $u\sim U(-0.25,0.25)$}\\ \hline\hline
\multicolumn{2}{c|}{$N=1000$} & Simulated & LPM & Ramp & Probit & Logit & Nonpar. \\
 &  & Truth* & (OLS) & (NLS) & (QMLE) & (QMLE) & (LL) \\ \hline
\multirow{2}{4em}{$APE_1$} &mean  & 0.2203  &0.0478  &0.2208  &0.2131  &0.2100  &0.1573\\
                           &sd    & 0.0063  &0.0100  &0.0163  &0.0149  &0.0154  &0.0289\\ \hline
\multirow{2}{4em}{$APE_2$} &mean  & -0.3251 &-0.2772 &-0.3252 &-0.3283 &-0.3271 &-0.3420\\
                           &sd    & 0.0083  &0.0263  &0.0213  &0.0203  &0.0205  &0.0233\\ \hline
\multicolumn{2}{c|}{$P(y=1)$}     & 0.8149 &   &  & &  &  \\
\multicolumn{2}{c|}{$P(0 \leq\widehat{y}\leq 1)$} &  & 0.9295  & 0.5483 &  1.0000 & 1.0000  & 0.8720\\ 
\multicolumn{2}{c|}{$P(-a\leq\mathbf{x\beta}\leq a)$} & 0.5507 &  &  &  &  &  \\ \hline 
\end{tabular}
\begin{tablenotes}
	\small
	\item *This column contains the empirical means and standard deviations of the sample average partial effects at the true parameter values.
	\end{tablenotes}
    \end{threeparttable}
\end{table}

\begin{table}[th]
    \centering
    \small
    \begin{threeparttable}
    \begin{tabular}
[c]{cc|cccccc}%
\multicolumn{8}{c}{Table 7. No interaction, $x_1$ asym., $x_2$ asym. binary, $u\sim U(-1,1)$}\\ \hline\hline
\multicolumn{2}{c|}{$N=1000$} & Simulated & LPM & Ramp & Probit & Logit & Nonpar. \\
 &  & Truth* & (OLS) & (NLS) & (QMLE) & (QMLE) & (LL) \\ \hline
\multirow{2}{4em}{$APE_1$} &mean  &0.0932	& 0.0477 &0.0939 &0.1083	  &0.1102	&0.0889 \\
                           &sd    &0.0008	& 0.0092 	&0.0095 &0.0095	  &0.0100	&0.0137 \\ \hline
\multirow{2}{4em}{$APE_2$} &mean  &-0.1379	&-0.1040   &-0.1391& -0.1428 &-0.1426&	-0.1412 \\
                           &sd    &0.0012	& 0.0318 	&0.0289 &0.0279	  &0.0280	&0.0296 \\ \hline
\multicolumn{2}{c|}{$P(y=1)$}     & 0.6453 & & & & &  \\
\multicolumn{2}{c|}{$P(0 \leq\widehat{y}\leq 1)$} &  & 0.9802  & 0.9317 &  1.0000 & 1.0000 &0.9803\\ 
\multicolumn{2}{c|}{$P(-a\leq\mathbf{x\beta}\leq a)$} & 0.9322 &  &  &  &  &  \\ \hline 
\end{tabular}
\begin{tablenotes}
	\small
	\item *This column contains the empirical means and standard deviations of the sample average partial effects at the true parameter values.
	\end{tablenotes}
    \end{threeparttable}
\end{table}

The findings in Table 7 are striking. Even though ${\small P}\left( -a\leq 
\mathbf{x\beta }\leq a\right) $ is high---around $0.93$---and the OLS fitted
values are very rarely outside the unit interval (only about 2 percent of the
time), the OLS estimators of the LPM are badly biased for the APEs and are
notably worse than other methods. And among the other estimators, Ramp/NLS has a smaller bias in terms of both APEs. In Table 6, it is the same case that Ramp/NLS continues to work relatively better than any other methods in terms of APEs, even though the fraction of fitted values of Ramp/NLS within the unit interval is only about 0.55. The results with an interaction term is similar and so are skipped for brevity.

\subsection{Additional Simulations}
Comparing Table 6 and Table 3, it seems like the symmetric distribution of $\mathbf{x}$ is the key condition for OLS to consistently estimate $APEs$. In Table 8, however, we give a counterexample where $\mathbf{x}$ is symmetrically distributed, but $x_1$ has a $Uniform(-10,10)$ distribution. Compared to the normal distribution of $x_1$ in Section 5.1, this distribution has higher variance and lacks a mode. Unlike before, OLS poorly approximates the APEs in this case. In addition to symmetry, therefore, the modality and higher moments of the covariates may also be important in determining the performance of OLS.

\begin{table}[th]
    \centering
    \small
    \begin{threeparttable}
    \begin{tabular}
[c]{cc|cccccc}%
\multicolumn{8}{c}{Table 8. No interaction, $x_1\sim U(-10,10)$, $x_2$ sym. binary, $u\sim U(-1,1)$}\\ \hline\hline
\multicolumn{2}{c|}{$N=1000$} & Simulated & LPM & Ramp & Probit & Logit & Nonpar. \\
 &  & Truth* & (OLS) & (NLS) & (QMLE) & (QMLE) & (LL) \\ \hline
\multirow{2}{4em}{$APE_1$} &mean  & 0.0501  &0.0682  &0.0501  &0.0500  &0.0498  &0.0512\\
                           &sd    & 0.0016  &0.0009  &0.0016  &0.0016  &0.0017  &0.0019\\ \hline
\multirow{2}{4em}{$APE_2$} &mean  & -0.0750 &-0.0750 &-0.0754 &-0.0754 &-0.0754 &-0.0752\\
                           &sd    & 0.0022  &0.0191  &0.0184  &0.0172  &0.0180  &0.0180\\ \hline
\multicolumn{2}{c|}{$P(y=1)$}     & 0.4801 &   &  &  &  &  \\
\multicolumn{2}{c|}{$P(0 \leq\widehat{y}\leq 1)$} &  & 0.7325 & 0.4960 & 1.0000 & 1.0000  & 0.9440\\ 
\multicolumn{2}{c|}{$P(-a\leq\mathbf{x\beta}\leq a)$} & 0.5006 &   &  &  &  &  \\ \hline 
\end{tabular}
\begin{tablenotes}
	\footnotesize
	\item *This column contains the empirical means and standard deviations of the sample average partial effects at the true parameter values. \\
	\end{tablenotes}
    \end{threeparttable}
\end{table}

An additional set of simulations are included in the Appendix which suggest our findings have more to do with the joint distribution of the explanatory variables than with the choice of distribution for the latent model error. Tables 11 corresponds to the DGPs of Tables 1-3, and Table 12 corresponds to the DGP of Tables 4-5, but the Appendix tables are based on standard normal error terms, corresponding to the probit model. As in main text, each of the estimators (including OLS) has small bias for the APEs in Tables 11-12. Table 13 in the Appendix corresponds to Tables 6-7 of the main text, but also with a normally distributed error. In this case, we find (as in the main text), OLS has larger bias for the APEs than do the nonlinear or nonparametric estimators.

\section{Mortgage Approval Probabilities and Race}
As an illustration of linear and nonlinear estimators for binary response models, we revisit the analysis of discrimination in mortgage lending decisions from \cite{hunter1996cultural}.\footnote{We use a version of the loan applications dataset provided by Mary Beth Walker for Wooldridge (2019).} The cultural affinity hypothesis posits that white loan officers may ``rely more heavily on basic objective loan application information in appraising the creditworthiness of minorities'' due to a lack of cultural familiarity. We compare linear and nonlinear estimates of the average effect of being white on the probability of loan approval, holding constant a number of loan, property, and borrower characteristics. Table 9 presents basic summary statistics for the dependent variable ``approve'' and 23 covariates. 

\begin{table}[th]
        \small
	\centering
	\begin{threeparttable}
	\begin{tabular}{llrrrr}
 \multicolumn{6}{c}{Table 9: Loan Approval Summary Statistics ($N=1989$)} \\
Variable & Description & Mean & SD & Skew. & Kurt. \\ \hline \hline
    approve & =1 if loan approved & 0.88 & 0.33 & -2.30 & 6.29\\ 
    white & =1 if white & 0.85 & 0.36 & -1.91 & 4.64\\ 
    loanamt & Loan amount \$1000s & 143.25 & 80.52 & 3.13 & 20.36\\ 
    suffolk & =1 if in Suffolk County & 0.15 & 0.36 & 1.91 & 4.66\\ 
    appinc & Applicant income \$1000s & 84.68 & 87.06 & 5.26 & 36.70\\ 
    unit & Number of units in property & 1.12 & 0.44 & 4.01 & 19.89\\ 
    married & =1 if applicant married & 0.66 & 0.47 & -0.67 & 1.45\\ 
    dep & Number of dependents & 0.77 & 1.10 & 1.47 & 5.33\\ 
    emp & Years employed in line of work & 0.21 & 1.00 & 6.69 & 50.57\\ 
    yjob & Years at this job & 0.45 & 1.12 & 5.32 & 36.18\\ 
    atotinc & Total monthly income & 5195.55 & 5269.06 & 6.36 & 65.34\\ 
    self & =1 if self employed & 0.13 & 0.34 & 2.21 & 5.89\\ 
    other & Other financing \$1000s & 2.37 & 28.23 & 26.80 & 886.84\\ 
    rep & Number of credit reports & 1.50 & 0.99 & 1.45 & 7.37\\ 
    pubrec & =1 if filed bankruptcy & 0.07 & 0.25 & 3.40 & 12.59\\ 
    hrat & Housing expense \% of total inc. & 24.79 & 7.12 & 0.25 & 6.74\\ 
    obrat & Other obligations \% of total inc. & 32.39 & 8.26 & 0.44 & 7.40\\ 
    cosign & =1 if there is a cosigner & 0.03 & 0.17 & 5.65 & 32.92\\ 
    sch & =1 if $>$ 12 years schooling & 0.77 & 0.42 & -1.29 & 2.68\\ 
    mortno & =1 if no mortgage history & 0.33 & 0.47 & 0.71 & 1.51\\ 
    mortlat1 & =1 if one or two late payments & 0.02 & 0.14 & 7.03 & 50.36\\ 
    mortlat2 & =1 if more than two late payments & 0.01 & 0.10 & 9.58 & 92.72\\ 
    chist & =0 if accounts are delinq. $\geq$60 days & 0.84 & 0.37 & -1.83 & 4.35\\ 
    loanprc & Loan amount / purchase price & 0.77 & 0.19 & 0.44 & 14.39\\ 
	\end{tabular}
	\end{threeparttable}
\end{table}

For our index model, we include interactions between ``white'' and all other explanatory variables to allow for the factors like loan amount and credit history to have a differential impact on approval probability by race. Let $w$ denote ``white'' and $\mathbf{z}$ be a vector including the 22 other covariates, so that $\mathbf{x}=\left\{1, \mathbf{z}, w, w\mathbf{z}\right\}$ and $\mathbf{\beta} = \left\{\mathbf{\beta}_0, \mathbf{\beta}_z, \beta_w, \mathbf{\beta}_{wz}\right\}$, where $\mathbf{\beta}_0$ is the intercept, $\mathbf{\beta}_z$ and $\beta_w$ are the coefficients on $\mathbf{z}$ and $w$, respectively, while $\mathbf{\beta}_{wz}$ is the coefficient on $w\mathbf{z}$.
Then the partial effects we average are formed by evaluating the difference in the probabilities evaluated at $w=1$ and $w=0$, respectively, as given below. \vspace{-3mm}
\begin{equation*}
    APE_w = E\left[G(\mathbf{\beta}_0+\beta_w + \mathbf{z}(\mathbf{\beta}_z+\mathbf{\beta}_{wz})) - G(\mathbf{\beta}_0+ \mathbf{z}\mathbf{\beta}_z)\right],  \vspace{-3mm}
\end{equation*}
where $G()$ is either the identity function (for the LPM estimated by OLS), the probit CDF, the logit CDF, or the ramp function. We also use the nonparametric kernel estimator from our simulations, treating ``white'' as discrete but not imposing a linear index or specific latent error distribution. 

Table 10 presents the results. Using the LPM estimated by OLS, about 18\% of observations have predicted probabilities outside the unit interval, so the H-O results clearly imply OLS is inconsistent for the slope parameters if the ramp model is correct. There is little reason to expect OLS will approximate this APE, either based on the theoretical results of \cite{Stoker1986} or our simulation study. Many of the explanatory variables are binary, and the continuous variables (e.g., income) tend to be skewed. For each variable, normality is strongly rejected by a Jarque-Bera test (a joint test of the skewness and kurtosis) with p-values well below 1\%. The model also includes interactions between the continuous variables and a binary variable. Using the LPM estimates, the APE for $white$ is $5.3$ percentage points and it is only marginally significant. Using the nonlinear parametric estimators, the APE are each a bit larger at about $7.0$ percentage points, and they are all significant at the 1\% level. Using the nonparametric estimator, the point estimate is quite a bit larger at 17.2 percentage points, but it is only marginally significant due to a very large standard error.

\begin{table}[htb!]
	\centering
        \small
	\begin{threeparttable}
	\begin{tabular}{lrrrrr}
       \multicolumn{6}{c}{Table 10: Estimates of the APE of ``White'' on Loan Approval ($N=1976$)} \\
	                         & LPM     & Ramp & Probit    & Logit    & Nonpar. \\
                          & (OLS) & (NLS) & (QMLE) & (QMLE) & (LL)\\\hline\hline
       Estimate                  & 0.0532    & 0.0706    & 0.0695    & 0.0712      & 0.1721  \\
       Robust SE                 & 0.0278    & 0.0227    & 0.0220    & 0.0219     & 0.0955 \\
       $P(0\leq \hat{y} \leq 1)$ & 0.8173    & 0.6027    & 1.0000    & 1.0000      & 0.6587 \\
       Mean Squared Error        & 0.1171    & 0.0839    & 0.0840    & 0.0837     & 0.1804 \\\hline
	\end{tabular}
	\begin{tablenotes}
	\small
	\item Note: There were only 1976 complete cases out of 1989 observations total. The nonparametric regression used only 1975 observations. All robust standard errors were computed using the sandwich forms and the delta method, save the nonparametric regression for which we used a nonparametric bootstrap with 500 replications.
	\end{tablenotes}
	\end{threeparttable}
\end{table}

Interestingly, OLS predicts only 18\% of observations with indexes outside the unit interval, whereas NLS predicts nearly 40\%, which follows the pattern of many of our simulations from the previous section and suggests trimming the sample once is not sufficient to consistently estimate the parameters or APEs under the piecewise linear model. Model selection by the minimum mean squared error favors logit, though the other nonlinear models are very similar.

\section{Implications for Empirical Research}
We\ have revisited the conclusions reached by \cite{Horrace2006}
concerning the ability of the linear projection parameters---consistently
estimated by OLS---to recover interesting parameters. We argue that H-O's
focus on the parameters in the underlying index model is misguided; instead,
one should focus on the APEs. Focusing on the APEs is hardly controversial,
as almost every study that employs any model nonlinear in the explanatory
variables reports estimated APEs.

Once the focus is on the APEs, a few useful conclusions emerge in an
expanded version of the H-O model, which allows for varying support in the
underlying uniform distribution. First, when the explanatory variables have
a multivariate normal distribution, the LP parameters are
identical to the population APEs under a general index model. Importantly, this is true even when the flat parts of the ramp function occur with high probability.
In this case, the LP parameters, $\gamma _{j}$, will be greatly attenuated
toward zero compared with the index parameters, $\beta _{j}$. The logit
and probit models, estimated by quasi-MLE (because the response
probabilities are misspecified), also approximate the APEs very well. Nonlinear least squares estimation of the ramp function is a new option, and we have shown the estimator is consistent for the best MSE approximation and asymptotically normal.

When the explanatory variables have asymmetric distributions, the conclusions
for OLS are not as sanguine---unless the support of $\mathbf{x\beta }$
is contained entirely in the support $\left[ -a,a\right] $ of the uniform
distribution of the the latent error. Some simulations show that
even if the probability of $\mathbf{x\beta }$ being in the unit interval is high (e.g. 93\% in Table 7), the LP\ parameters are not very
close to the true APEs. Especially when the support $\left[ -a,a\right] $ is narrow, the logit and
probit approximations to the APEs can be notably better than those for OLS.

To summarize, in evaluating different strategies, we need to make sure we have carefully defined the population quantities of interest, and then we make proper comparisons across different approaches.
OLS estimation of the LPM has good finite sample properties for the APE in many cases when the covariates are symmetrically distributed. Probit, Logit, and the ramp model continue to have good finite sample properties for estimating the APEs when the covariates are asymmetrically distributed. Using the fraction of estimated response probabilities in $\left[ 0,1\right]$ is neither necessary nor sufficient for good performance of OLS. In the $a=1/4$ case with
interactions, the estimated probability is almost $0.95$ but OLS has a severe bias toward zero for the APEs. By contrast, the three nonlinear models show very little bias. A nonlinear model, of course, offers other advantages over the LPM, such as more realistic response probabilities and nonconstant partial effects. However, when the APEs are of interest, OLS is more widely applicable than a simple reading of H-O might suggest.

The conclusions drawn here are easily extended to the case where $y$ is a
fractional response, where the limit values zero and one can occur with
positive probability. In particular, Stoker (1986) can be applied to $%
E\left( y|\mathbf{x}\right) $. If this conditional mean follows the 
same ramp function, the qualitative conclusions obtained
in the binary case will remain.

\bibliographystyle{apalike}
\bibliography{lpm.bib}

\section*{Appendix}

\subsection*{Proof of Theorem 2}
\noindent$Proof.$ We will obtain the asymptotic normality of the NLS estimator by applying Theorem 7.1 of Newey and McFadden (1994). Condition (i) and (ii) of Theorem 7.1 follows from our assumptions. As we discussed in the main context, condition (iii) is satisfied as long as $\mathbf{x}$ contains a continuous variables $x_j$ with nonzero $\beta_{jo}$ so that $P(\mathbf{x}_i\mathbf{\beta}_o=0 \text{ or }\mathbf{x}_i\mathbf{\beta}_o=1) = 0$. 

For condition (iv), notice that the first derivative of the object function is well defined at $\mathbf{\beta}_o$ with probability 1:  \vspace{-3mm}
\begin{align*}
    D_N(\mathbf{\beta}_o) = \nabla_{\beta}Q_N(\mathbf{\beta}_o) = \frac{1}{N} \sum_{i=1}^N \mathbf{x}_i' (y_i-\mathbf{x}_i\mathbf{\beta}_o) 1\{\mathbf{x}_i\mathbf{\beta}_o\in(0,1)\}= \frac{1}{N} \sum_{i=1}^N \mathbf{x}_i'u_i1\{\mathbf{x}_i\mathbf{\beta}_o\in(0,1)\}, \vspace{-3mm}
\end{align*}
where $u_i = y_i - R(\mathbf{x}_i\mathbf{\beta}_o)$. Since $E\Vert \mathbf{x}_i'u_i\Vert1\{\mathbf{x}_i\mathbf{\beta}_o\in(0,1)\}<\infty$ under the assumption $E\Vert x\Vert^2<\infty$, the vector Lindberg-Levy CLT applies: \vspace{-3mm}
\begin{align*}
    \sqrt{N}D_N(\mathbf{\beta}_o) \stackrel{d}{\to} \mathbb{N}\left(0,\mathbf{\Omega}(\mathbf{\beta}_o)\right), \vspace{-3mm}
\end{align*}
giving condition (iv). Lastly, for condition (v), following Newey and McFadden (1994), we can rewrite  \vspace{-3mm} 
\begin{align*}
    &\sqrt{N}[Q_N(\mathbf{\beta})-Q_N(\mathbf{\beta}_o)] \\
    = &\sqrt{N}\left[D_N(\mathbf{\beta}_o)(\mathbf{\beta}-\mathbf{\beta}_o)+Q(\mathbf{\beta})-Q(\mathbf{\beta}_o)\right] +\Vert\mathbf{\beta}-\mathbf{\beta}_o\Vert M_N(\mathbf{\beta}), \vspace{-3mm}
\end{align*}
where $M_N(\mathbf{\beta})$ is the remainder term, defined as: \vspace{-3mm}
\begin{align*}
    M_N(\mathbf{\beta})=\frac{\sqrt{N}\left[Q_N(\mathbf{\beta})-Q_N(\mathbf{\beta}_o) - D_N'(\mathbf{\beta}_o)(\mathbf{\beta}-\mathbf{\beta}_o)-\left(Q(\mathbf{\beta})-Q(\mathbf{\beta}_o)\right)\right]}{\Vert\mathbf{\beta}-\mathbf{\beta}_o\Vert}. \vspace{-3mm}
\end{align*}

Since $D_N(\mathbf{\beta}_o)$ is the gradient of $Q_N(\mathbf{\beta})$ at $\mathbf{\beta}_o$, $Q_N(\mathbf{\beta})-Q_N(\mathbf{\beta}_o)-D_N(\mathbf{\beta}_o)(\mathbf{\beta}-\mathbf{\beta}_o)$ goes to zero faster than $\Vert\mathbf{\beta}-\mathbf{\beta}_o\Vert$ as $\mathbf{\beta}$ goes to $\mathbf{\beta}_o$, by the definition of the gradient. Similarly, due to $\nabla_{\beta}Q(\mathbf{\beta}_o) = E\left(s_i(\mathbf{\beta}_o)\right)=0$, $Q(\mathbf{\beta})-Q(\mathbf{\beta}_o)$ goes to 0 faster than $\Vert \mathbf{\beta}-\mathbf{\beta}_o\Vert$ as $\mathbf{\beta}$ goes to $\mathbf{\beta}_o$. Under the moment conditions, we can easily show $Q(\mathbf{\beta})-Q(\mathbf{\beta}_o) \to 0$ in probability and so $\sqrt{N}[Q_N(\mathbf{\beta})-Q(\mathbf{\beta})]$ is bounded in probability for each $\beta$. Also note that $\sqrt{N}D_N(\mathbf{\beta}_o)$ is bounded in probability due to the asymptotic normality. Since the numerator is bounded in probability and converges to zero faster than the denominator, we conclude that for any $\varepsilon_N>0$, $\lim_{N\to\infty} sup_{\Vert \mathbf{\beta}-\mathbf{\beta}_o\Vert<\varepsilon_N} M_N(\mathbf{\beta}) \to 0$ in probability, which implies condition (v).
\qedsymbol{}
\bigskip
\subsection*{Proof of Theorem 3}
\noindent$Proof.$ Consider $\mathbf{\Omega}(\mathbf{\hat{\beta}})$: \vspace{-3mm}
\begin{align*}
    \mathbf{\Omega}(\mathbf{\hat{\beta}}) = & \frac{1}{N} \sum_{i=1}^N \mathbf{x}_i'\mathbf{x}_i \left( y_i - R(\mathbf{x}_i\hat{\mathbf{\beta}})\right)^2 1\{\mathbf{x}_i\mathbf{\hat{\beta}}\in(0,1)\} \\  
    \equiv &  \frac{1}{N} \sum_{i=1}^N a(\mathbf{x_i},\hat{\mathbf{\beta}})  \vspace{-3mm}
\end{align*}
Note that $E| y_i - R(\mathbf{x}_i\mathbf{\beta})|^4 \leq 1$ for any $\mathbf{\beta} \in \mathcal{B}$ since both $y_i$ and $R(.)$ are naturally bounded in $[0,1]$ with probability 1. Then, we have \vspace{-3mm}
\begin{align*}
    E \sup_{\mathbf{\beta}\in \mathcal{B}} \Vert a(x,\mathbf{\beta})\Vert \leq (E\left\Vert \mathbf{x}_i\right\Vert^4 E| y_i - R(\mathbf{x}_i\mathbf{\beta})|^4)^{1/2}<\infty,  \vspace{-3mm}
\end{align*}
where the first inequality follows from H\"{o}lder's inequality. Also note that $a(\mathbf{x_i},\mathbf{\beta})$ is continuous at $\mathbf{\beta}_o$ with probability one given that $P(\mathbf{x}_i\mathbf{\beta}_o = 0) = P(\mathbf{x}_i\mathbf{\beta}_o = 1) =0$. Then, we can apply Lemma 4.3 of \cite{Newey1994}:  \vspace{-3mm}
\begin{align*}
     \mathbf{\Omega}(\mathbf{\hat{\beta}}) = \frac{1}{N} \sum_{i=1}^N a(\mathbf{x_i},\hat{\mathbf{\beta}})  \stackrel{p}{\to} E(a(\mathbf{x_i}, \mathbf{\beta}_o)) =\mathbf{\Omega}(\mathbf{\beta}_o). \vspace{-3mm}
\end{align*}
Similarly, Lemma 4.3 also applies to $\mathbf{A}_N(\mathbf{\hat{\beta}}) = \frac{1}{N} \sum_{i=1}^N  \mathbf{x}_i'\mathbf{x}_i 1\{\mathbf{x}_i\mathbf{\hat{\beta}}\in(0,1)\}$: \vspace{-3mm}
\begin{align*}
   \plim_{N\to\infty} \frac{1}{N} \sum_{i=1}^N 1\{\mathbf{x}_i\mathbf{\hat{\beta}}\in(0,1)\} \mathbf{x}_i'\mathbf{x}_i = \mathbf{A}(\mathbf{\beta}_o). \vspace{-3mm}
\end{align*}
So, we conclude that \vspace{-3mm}
\begin{align*}
     \mathbf{\hat{V}} = \mathbf{A}_N(\hat{\mathbf{\beta}})^{-1} \mathbf{\Omega}_N(\hat{\mathbf{\beta}}) \mathbf{A}_N(\hat{\mathbf{\beta}})^{-1} \stackrel{p}{\to} \mathbf{A}(\mathbf{\beta}_o)^{-1} \mathbf{\Omega}(\mathbf{\beta}_o)\mathbf{A}(\mathbf{\beta}_o)^{-1}. \, \, \qedsymbol{} \vspace{-3mm}
\end{align*}

\textbf{\large Additional Simulations} 
\begin{table}[th] \vspace{-1.5mm}
    \small
    \centering
    \begin{threeparttable}
    \begin{tabular}
[c]{cc|cccccc}%
\multicolumn{8}{c}{Table 11. No interaction, $x_1$ sym, $x_2$ sym. binary, $u\sim N(0,1)$}\\ \hline\hline
\multicolumn{2}{c|}{$N=1000$} & Simulated & LPM & Ramp & Probit & Logit & Nonpar. \\
 &  & Truth* & (OLS) & (NLS) & (QMLE) & (QMLE) & (LL) \\ \hline
\multirow{2}{4em}{$APE_1$} &mean  & 0.0782  &0.0774  &0.0774  &0.0774  &0.0774  &0.0776\\
                           &sd    & 0.0001  &0.0171  &0.0171  &0.0171  &0.0171  &0.0175\\ \hline
\multirow{2}{4em}{$APE_2$} &mean  & -0.1168 &-0.1160 &-0.1160 &-0.1154 &-0.1152 &-0.1160\\
                           &sd    & 0.0001  &0.0339  &0.0339  &0.0336  &0.0335  &0.0344\\ \hline
\multicolumn{2}{c|}{$P(y=1)$}     & 0.4805 &   &  &  &  &  \\
\multicolumn{2}{c|}{$P(0 \leq\widehat{y}\leq 1)$} &  & 1.0000 & 1.0000 & 1.0000 & 1.0000  & 0.9983\\\hline 
\end{tabular}
\begin{tablenotes}
	\footnotesize
	\item *This column contains the empirical means and standard deviations of the sample average partial effects at the true parameter values. \\
	\end{tablenotes}
    \end{threeparttable}
    
    \begin{threeparttable}
    \begin{tabular}
[c]{c|c|cccccc}%
\multicolumn{8}{c}{Table 12. With interaction, $x_1$ sym, $x_2$ sym. binary, $u\sim N(0,1)$}\\ \hline\hline
\multicolumn{2}{c|}{$N=1000$} & Simulated & LPM & Ramp & Probit & Logit & Nonpar. \\
 &  & Truth* & (OLS) & (NLS) & (QMLE) & (QMLE) & (LL) \\ \hline
\multirow{2}{4em}{$APE_1$} &mean  & 0.0199 & 0.0191  &0.0191  &0.0191  &0.0191  &0.0193\\
                           &sd    & 0.0019 & 0.0171  &0.0170  &0.0171  &0.0171  &0.0176\\ \hline
\multirow{2}{4em}{$APE_2$} &mean  & -0.1179& -0.1177 &-0.1177 &-0.1173 &-0.1172 &-0.1178\\
                           &sd    & 0.0037 & 0.0337  &0.0337  &0.0336  &0.0336  &0.0347\\ \hline
\multicolumn{2}{c|}{$P(y=1)$}     & 0.4563 &   &  &  &  &  \\
\multicolumn{2}{c|}{$P(0 \leq\widehat{y}\leq 1)$} &  & 0.9999  & 0.9999 & 1.0000 & 1.0000  & 0.9983\\\hline 
\end{tabular}
\begin{tablenotes}
	\footnotesize
	\item *This column contains the empirical means and standard deviations of the sample average partial effects at the true parameter values. \\
	\end{tablenotes}
    \end{threeparttable}
    
    \begin{threeparttable}
    \begin{tabular}
[c]{cc|cccccc}%
\multicolumn{8}{c}{Table 13. No interaction, $x_1$ asym, $x_2$ asym. binary, $u\sim N(0,1)$}\\ \hline\hline
\multicolumn{2}{c|}{$N=1000$} & Simulated & LPM & Ramp & Probit & Logit & Nonpar. \\
 &  & Truth* & (OLS) & (NLS) & (QMLE) & (QMLE) & (LL) \\ \hline
\multirow{2}{4em}{$APE_1$} &mean  & 0.0722 & 0.0410  &0.0667  &0.0728  &0.0753  &0.0672\\
                           &sd    & 0.0005 & 0.0077  &0.0101  &0.0094  &0.0099  &0.0134\\ \hline
\multirow{2}{4em}{$APE_2$} &mean  & -0.1070& -0.0840 &-0.1064 &-0.1075 &-0.1083 &-0.1092\\
                           &sd    & 0.0007 & 0.0322  &0.0305  &0.0296  &0.0294  &0.0315\\ \hline
\multicolumn{2}{c|}{$P(y=1)$}     & 0.6143 &   &  &  &  &  \\
\multicolumn{2}{c|}{$P(0 \leq\widehat{y}\leq 1)$} &&0.9873&0.9659&1.0000&1.0000&0.9844\\\hline 
\end{tabular}
\begin{tablenotes}
	\footnotesize
	\item *This column contains the empirical means and standard deviations of the sample average partial effects at the true parameter values. 
	\end{tablenotes}
    \end{threeparttable}
\end{table} 
\end{document}